\documentclass{article}
\usepackage{amssymb}
\usepackage{amsfonts}
\usepackage{amsmath}

\newtheorem{theorem}{Theorem}

\newtheorem{claim}[theorem]{Claim}

\newtheorem{conjecture}[theorem]{Conjecture}
\newtheorem{corollary}[theorem]{Corollary}

\newtheorem{proposition}[theorem]{Proposition}

\newenvironment{proof}[1][Proof]{\noindent\textbf{#1.} }{\ \rule{0.5em}{0.5em}}
\input{tcilatex}

\begin{document}

\title{$5$-cycles and the Petersen graph}
\author{M. DeVos$^{\perp }$, V. V. Mkrtchyan\dag \ddag \thanks{%
The author is supported by a grant of Armenian National Science and
Education Fund}, S. S. Petrosyan\dag , \\
%EndAName
$^{\perp }$Department of Mathematics, Simon Fraser University, Canada\\
\dag Department of Informatics and Applied Mathematics,\\
Yerevan State University, Yerevan, 0025, Armenia\\
\ddag Institute for Informatics and Automation Problems, NAS\\
RA, 0014, Armenia \\
email: mdevos@sfu.ca\\
vahanmkrtchyan2002@\{ysu.am, ipia.sci.am, yahoo.com\}\\
samvelpetrosyan@gmail.com \textit{\text{ }}}
\maketitle

\begin{abstract}
We show that if $G$ is a connected bridgeless cubic graph whose every $2$%
-factor is comprised of cycles of length five then $G$ is the Petersen graph.
\end{abstract}

\bigskip

"The Petersen graph is an obstruction to many properties in graph theory,
and often is, or conjectured to be, the only obstruction". This phrase is
taken from one of the series of papers by Robertson, Sanders, Seymour and
Thomas that is devoted to the proof of prominent Tutte conjecture- a
conjecture which states that if the Petersen graph is not a minor of a
bridgeless cubic graph $G$ then $G$ is $3$-edge-colorable, and which in its
turn is a particular case of a much more general conjecture of Tutte stating
that every bridgeless graph $G$ has a nowhere zero $4$-flow unless the
Petersen graph is not a minor of $G$.

Another result that stresses the exceptional role of the Petersen graph is
proved by Alspach et al. in \cite{Alspach}. The following striking
conjecture of Jaeger states that everything related to the colorings of
bridgeless cubic graphs can be reduced to that of the Petersen graph, more
specifically,

\begin{conjecture}
Petersen coloring conjecture of Jaeger \cite{JaegerFlows}: the edges of
every bridgeless cubic graph $G$ can be mapped into the edges of the
Petersen graph in such a way that any three mutually incident edges of $G$
are mapped to three mutually incident edges of the Petersen graph.
\end{conjecture}

The Petersen graph is so important in graph theory that even an entire book
is written about it \cite{PetersenBook}.

It is known that for every bridgeless cubic graph $G$ and its two edges $e$
and $f$ we can always find a $2$-factor of $G$ containing these two edges.
Recently, Jackson and Yoshimoto in \cite{Jackson} observed that in any
bridgeless cubic graph without multiple edges we can always find a
triangle-free $2$-factor. An earlier result by Rosenfeld \cite{Rosenfeld}
also worths to be mentioned here, which states that there are infinitely
many $3$-connected cubic graphs whose every $2$-factor contains a cycle of
length no more five.

Sometimes it is convinient to have a $2$-factor $F$ of a bridgeless cubic
graph $G$ such that not all cycles of $F$ are $5$-cycles ($=$cycles of
length five) \cite{Tractatus}. The main reason why we are interested in $5$%
-cycles is the following: it is known that it is the odd cycles of a graph
(particularly, a bridgeless cubic graph) that prevent it to have a $3$%
-edge-coloring. Fortunately, the triangles can be overcome easily. This is
due to the operation of the contraction of the triangles, which preserves
the cubicness and bridgelessness of a graph. Therefore, we need a technique
to cope with odd cycles of length at least five, and particularly, the $5$%
-cycles. The main result of this paper states that unless $G$ is the
Petersen graph in every connected bridgeless cubic graph $G$ we can always
find a $2$-factor $F$ that contains a cycle which is not a $5$-cycle.

\bigskip

We consider finite, undirected graphs without loops. Graphs may contain
multiple edges. We follow \cite{Lov, West} for the terminology.

\begin{theorem}
\label{Tutte}(Tutte, \cite{Tutte}): A graph $G$ contains a perfect matching
if and only if for every $S\subseteq V(G)$ $o(G-S)\leq \left\vert
S\right\vert $, where $o(H)$ denotes the number of odd components of a graph 
$H$.
\end{theorem}

\begin{corollary}
\label{ContaingEdge}If $G$ is a bridgeless cubic graph, then for every its
edge $e$ there is a perfect matching $F$ with $e\in F$.
\end{corollary}

This corollary immediately implies

\begin{corollary}
\label{MissingEdge}If $G$ is a bridgeless cubic graph, then for every its
edge $e$ there is a perfect matching $F$ with $e\notin F$.
\end{corollary}

We will also need the following property of the Petersen graph:

\begin{proposition}
\label{PetersenProperty}The Petersen graph is the unique cubic graph of
girth five on ten vertices.
\end{proposition}

We are ready to state the main result of the paper:

\begin{theorem}
If $G$ is a connected bridgeless cubic graph whose every $2$-factor is
comprised of $5$-cycles then $G$ is the Petersen graph.
\end{theorem}

\begin{proof}
First of all note that $G$ is not $3$-edge-colorable, thus every $2$-factor
of $G$ contains at least two odd cycles.

\begin{claim}
\label{Removing2-cycles}$G$ does not have a cycle of length two.
\end{claim}

\begin{proof}
Suppose that $G$ contains a cycle $C$ of length two. Let $u$ and $v$ be the
vertices of $C$, and let $u^{\prime },v^{\prime }$ be the other ($\neq
v,\neq u$) neighbours of $u$ and $v$, respectively. Note that since $G$ is
bridgeless, we have $u^{\prime }\neq v^{\prime }$. Now let $F$ be a perfect
matching of $G$ containing the edge $(u,u^{\prime })$ (corollary \ref%
{ContaingEdge}). Clearly, $(v,v^{\prime })\in F$. Consider the complementary 
$2$-factor of $F$. Note that $C$ is a cycle in this $2$-factor contradicting
the condition of the theorem.
\end{proof}

\begin{claim}
\label{RemovingAdjacentTriangles}$G$ does not have two triangles sharing an
edge.
\end{claim}

\begin{proof}
Let $u,v,w$ and $u^{\prime },v,w$ be two triangles of $G$ which share the
edge $\left( v,w\right) $. Clearly, $(u,u^{\prime })\notin E(G)$, as $G$ is
not $3$-edge-colorable. Let $u_{1}$ and $u_{1}^{\prime }$ be the other ($%
\neq v,w$) neighbours of $u$ and $u^{\prime }$, respectively. Note that
since $G$ is bridgeless we have $u_{1}\neq u_{1}^{\prime }$. Consider a
perfect matching $F$ with $\left( v,w\right) \in F$ (corollary \ref%
{ContaingEdge}). Clearly, $(u,u_{1}),(u_{1},u_{1}^{\prime })\in F$. Note
that the complementary $2$-factor of $F$ contains the $4$-cycle on vertices $%
u,v,u^{\prime },w$ contradicting the condition of theorem.
\end{proof}

\begin{claim}
\label{RemovingAdjacentSquareTriangles}$G$ does not have a square and a
triangle sharing an edge.
\end{claim}

\begin{proof}
Suppose, on the contrary, that $G$ contains a square $(u,v)$, $(v,x)$, $%
(x,w) $, $(w,u)$ and a triangle $(v,y),(y,x),(x,v)$ which share the edge $%
(x,v)$. Due to claim \ref{RemovingAdjacentTriangles}, $(u,x)\notin E(G)$, $%
(v,w)\notin E(G)$. Now, let $F$ be a perfect matching of $G$ containing the
edge $(u,w)$ (corollary \ref{ContaingEdge}). Clearly, $(v,x)\in F$ and there
is a vertex $y^{\prime }\notin \{u,v,x,w,y\}$ such that $(y,y^{\prime })\in
E(G)$. Now, consider a path $u,(u,v),v,(v,y),y,(y,x),x,(x,w),w$ of length
four. The path lies on a cycle $C$ of the complementary $2$-factor of $F$.
Due to claim \ref{Removing2-cycles} there is only one edge connecting $u$
and $w$. Thus the length of $C$ is at least six contradicting the condition
of theorem.
\end{proof}

\begin{claim}
\label{RemovingTriangles} $G$ does not have a triangle.
\end{claim}

\begin{proof}
On the opposite assumption, consider a triangle $C$ on vertices $x,y,z$ of $%
G $. Since $G$ is bridgeless we imply that there are vertices $x^{\prime
},y^{\prime },z^{\prime }$ adjacent to $x,y,z$, respectively, that do not
lie on $C$. Now, consider a perfect matching $F$ of $G$ containing the edge $%
(x,y)$ (corollary \ref{ContaingEdge}). Clearly, $(z,z^{\prime })\in F$. Note
that the path $y^{\prime },(y^{\prime },y),y,(y,z),z,(z,x),x,(x,x^{\prime
}),x^{\prime }$ of length four lies on some cycle $C^{\prime }$ of the
complementary $2$-factor of $F$. Claim \ref{RemovingAdjacentTriangles}
implies that $(y^{\prime },x^{\prime })\notin E(G)$ thus the length of $%
C^{\prime }$ is at least six contradicting the condition of theorem.
\end{proof}

\begin{claim}
\label{RemovingSquares}$G$ does not have a square, and girth of $G$ is five.
\end{claim}

\begin{proof}
Assume $G$ to contain a square $C=(u,v),(v,w),(w,z),(z,u)$. Claim \ref%
{RemovingTriangles} implies that $(u,w)\notin E(G)$, $(v,z)\notin E(G)$. Let 
$u_{1},v_{1},w_{1},z_{1}$ be the vertices of $G$ that are adjacent to $%
u,v,w,z$, respectively and do not lie on $C$. Consider a perfect matching $F$
of $G$ containing the edge $(u,u_{1})$ (corollary \ref{ContaingEdge}).
Clearly, $\{(v,v_{1}),(w,w_{1}),(z,z_{1})\}\nsubseteq F$, as if it were true
then the complementary $2$-factor of $F$ would have contained $C$ as a
cycle, which contradicts the condition of theorem. Thus 
\begin{equation*}
\left\vert F\cap \{(v,v_{1}),(w,w_{1}),(z,z_{1})\}\right\vert =1
\end{equation*}
Without loss of generality, we may assume that $(v,v_{1})\in F$. Note that $%
(w,z)\in F$. Now, consider the cycle $C_{F}$ in the complementary $2$-factor
of $F$, which contains the path $%
z_{1},(z_{1},z),z,(z,u),u,(u,v),v,(v,w),w,(w,w_{1}),w_{1}$. Due to claim \ref%
{RemovingTriangles} $z_{1}\neq w_{1}$ thus the length of $C_{F}$ is at least
six contradicting the condition of theorem. Thus, $G$ cannot contain a
square, too, therefore its girth is five.
\end{proof}

\begin{claim}
\label{Removing2Cuts}$G$ is $3$-edge-connected.
\end{claim}

\begin{proof}
Suppose, for a contradiction, that $G$ is only $2$-edge-connected, and let $%
(u,v),(u^{\prime },v^{\prime })$ be two edges which form a $2$-edge cut so
that $u$ and $u^{\prime }$ are in the same component of $G\backslash
\{(u,v),(u^{\prime },v^{\prime })\}$. Now, there must exist a perfect
matching not using $(u,v)$ (corollary \ref{MissingEdge}), so the
complementary $2$-factor must contain a $5$-cycle which uses both the edges $%
(u,v)$ and $(u^{\prime },v^{\prime })$. It follows that either $(u,u^{\prime
})\in E(G)$ or $(v,v^{\prime })\in E(G)$. Without loss of generality, we may
assume that $(v,v^{\prime })\in E(G)$. Let $w$ be the neighbor of $v$ other
than $u,v^{\prime }$, and let $w^{\prime }$ be the neighbor of $v^{\prime }$
other than $u^{\prime },v$. Now, there exists a perfect matching containing
the edge $(v,v^{\prime })$ (corollary \ref{ContaingEdge}), and the
complementary $2$-factor must contain a $5$-cycle which uses all of the
edges $(u,v),(v,w),(u^{\prime },v^{\prime }),(v^{\prime },w^{\prime })$. It
follows that either $u=u^{\prime }$ or $v=v^{\prime }$, but either of these
contradicts the fact that $G$ is bridgeless. This contradiction shows that $%
G $ is $3$-edge connected.
\end{proof}

\begin{claim}
\label{CyclicallyFourConnected} Every $3$-edge-cut of $G$ consists of three
edges incident to a common vertex.
\end{claim}

\begin{proof}
Let $(U,\bar{U})=\{(u_{1},v_{1}),(u_{2},v_{2}),(u_{3},v_{3})\}$ be a $3$-cut
of $G$ and suppose that $\left\{ u_{1},u_{2},u_{3}\right\} \subseteq U$, $%
\left\{ v_{1},v_{2},v_{3}\right\} \subseteq \bar{U}$. We claim that either $%
u_{1}=u_{2}=u_{3}$ or $v_{1}=v_{2}=v_{3}$. Before showing this let us show
that there is no edge connecting $u_{i}$ and $u_{j}$ or $v_{i}$ and $v_{j}$, 
$1\leq i<j\leq 3$. On the opposite assumption, suppose that $\left(
v_{1},v_{2}\right) \in E(G)$. Let $v_{1}^{\prime }$ and $v_{2}^{\prime }$ be
the neighbours of $v_{1}$ and $v_{2}$, respectively, that are different from 
$u_{1},v_{2}$ and $u_{2},v_{1}$. Clearly, $v_{1}^{\prime },v_{2}^{\prime
}\in \bar{U}$ and claim \ref{RemovingTriangles} implies that $v_{1}^{\prime
}\neq v_{2}^{\prime }$. Consider a perfect matching $F_{1,2}$ of $G$
containing the edge $\left( v_{1},v_{2}\right) $ (corollary \ref%
{ContaingEdge}). Since $\left\vert U\right\vert $ is odd ($(U,\bar{U})$ is
an odd cut), we have $(u_{3},v_{3})\in F_{1,2}$. Thus, the complementary $2$%
-factor of $F_{1,2}$ must contain a $5$-cycle containing the edges $%
(u_{1},v_{1}),(v_{1},v_{1}^{\prime }),(u_{2},v_{2}),(v_{2},v_{2}^{\prime })$%
. Since $v_{1}^{\prime }\neq v_{2}^{\prime }$ we have $u_{1}=u_{2}$. Note
that$\ u_{1},v_{1},v_{2}$ forms a triangle contradicting claim \ref%
{RemovingTriangles}. Thus $\left( v_{1},v_{2}\right) \notin E(G)$.
Similarly, the absence of the other edges can be shown. Now, let us turn to
the proof of claim \ref{CyclicallyFourConnected}. Let $F$ be a perfect
matching missing $(u_{1},v_{1})$ (corollary \ref{MissingEdge}). Since $%
\left\vert U\right\vert $ is odd, we imply that $F$ contains one of $%
(u_{2},v_{2}),(u_{3},v_{3})$ and misses the other one. Without loss of
generality, we may assume that $(u_{3},v_{3})\in F$, $(u_{2},v_{2})\notin F$%
. Note that there should be a $5$-cycle containing both the edges $%
(u_{2},v_{2})$ and $(u_{3},v_{3})$. As $(u_{1},u_{2})\notin F$, $\left(
v_{1},v_{2}\right) \notin F$, we imply that either $u_{1}=u_{2}$ or $%
v_{1}=v_{2}$. Again, we can assume that $u_{1}=u_{2}$. Let us show that $%
u_{1}=u_{3}$, too. Suppose that $u_{1}\neq u_{3}$. Let $w$ be a vertex from $%
U$ adjacent to the vertex $u_{1}=u_{2}$, and let $U^{\prime }=U\backslash
\{u_{1}\}$. Note that $(U^{\prime },\bar{U}^{\prime
})=\{(u_{1},w),(u_{3},v_{3})\}$ is a $2$-edge-cut of $G$ contradicting the
choice of claim \ref{Removing2Cuts}. Thus $u_{1}=u_{3}$ and we are done.
\end{proof}

\begin{claim}
\label{ThreepathinPerfects}If $u,v,w,x\in V(G)$ and $(u,v),(v,w),(w,x)\in
E(G)$ then there is a perfect matching of $G$ containing both $(u,v)$ and $%
(w,x)$.
\end{claim}

\begin{proof}
Suppose, for a contradiction, that the statement does not hold. Then $%
G^{\prime }=G\backslash \{u,v,w,x\}$ has no perfect matching, so by theorem %
\ref{Tutte} there exists a subset of vertices $Y\subseteq V(G^{\prime })$ so
that $G^{\prime }\backslash Y$ has more than $\left\vert Y\right\vert $ odd
components. Let $I$ be the set of isolated vertices in $G^{\prime
}\backslash Y$, let $O$ be the set of odd components of $G^{\prime
}\backslash Y$ with at least three vertices, and let $E$ be the set of even
components of $G^{\prime }\backslash Y$ . We know that $\left\vert
Y\right\vert <\left\vert I\right\vert +\left\vert O\right\vert $ by
assumption, but in fact $\left\vert Y\right\vert +2\leq \left\vert
I\right\vert +\left\vert O\right\vert $ since $\left\vert Y\right\vert
-\left\vert I\right\vert -\left\vert O\right\vert $ must be an even number
(as $\left\vert V(G^{\prime })\right\vert $ is even). Now, let $Y^{+}=Y\cup
\{u,v,w,x\}$ and let $C$ be the edge cut which separates $Y^{+}$ from $%
V(G)\backslash Y^{+}$. It follows from our construction that $\left\vert
C\right\vert \leq 3\left\vert Y\right\vert +6$ since every vertex in $Y$ can
contribute at most three edges to $C$, and there are at most six edges in $C$
with one of $u,v,w,x$ as endpoint. On the other hand, claim \ref%
{CyclicallyFourConnected} implies that every component in $O\cup E$ must
contribute at least four edges to $C$, and every vertex in $I$ contributes
exactly three edges to $C$, so 
\begin{equation*}
\left\vert C\right\vert \geq 3\left\vert I\right\vert +4\left\vert
O\right\vert +4\left\vert E\right\vert \geq 3(\left\vert Y\right\vert
+2)+\left\vert O\right\vert +4\left\vert E\right\vert
\end{equation*}%
It follows from this that $O=E=\varnothing $, and that every vertex in $Y$
must have all three incident edges in $C$. Thus $G\backslash
\{(u,v),(v,w),(w,x)\}$ is a bipartite graph. Now, there exists a perfect
matching of $G$ which contains the edge $(u,v)$, and every odd cycle in the
complementary $2$-factor must contain $(v,w)$ and $(w,x)$, so the
complementary $2$-factor cannot have two odd cycles - giving us a
contradiction.
\end{proof}

Now we are ready to complete the proof of the theorem. Claim \ref%
{ThreepathinPerfects} implies that every $3$-edge path must be contained in
a cycle of length five, and it follows from this that every $2$-edge path of
is contained in at least two $5$-cycles. Let $u$ be a vertex of $G$, let $%
v,w,x$ be the neighbors of $u$, and assume that the neighbors of $v,w,x$ are 
$\{u,v_{1},v_{2}\}$, $\{u,w_{1},w_{2}\}$, and $\{u,x_{1},x_{2}\}$,
respectively. It follows from the fact that $G$ has girth five that all of
these vertices we have named are distinct. Since the $2$-edge path with
edges $(v,u),(u,w)$ is contained in two cycles of length five, there must be
at least two edges between $\{v_{1},v_{2}\}$ and $\{w_{1},w_{2}\}$.
Similarly, there are at least two edges between $\{w_{1},w_{2}\}$ and $%
\{x_{1},x_{2}\}$, and between $\{x_{1},x_{2}\}$ and $\{v_{1},v_{2}\}$. As $G$
is connected we imply that $V(G)=%
\{u,v,w,x,v_{1},v_{2},w_{1},w_{2},x_{1},x_{2}\}$, and there are exactly two
edges between $\{v_{1},v_{2}\}$ and $\{w_{1},w_{2}\}$, $\{w_{1},w_{2}\}$ and 
$\{x_{1},x_{2}\}$, $\{x_{1},x_{2}\}$ and $\{v_{1},v_{2}\}$. Proposition \ref%
{PetersenProperty} implies that $G$ is isomorphic to the Petersen graph.
\end{proof}

\end{document}